\documentclass[aps,prb,superscriptaddress,showpacs,reprint]{revtex4-1}

\usepackage[utf8]{inputenc}
\usepackage{amsmath,amsfonts,amssymb}
\usepackage{pifont}
\usepackage{rotating}
\usepackage{enumerate}
\usepackage{filecontents}
\usepackage{float}
\usepackage{graphicx}    
\usepackage{epstopdf}
\usepackage[table,dvipsnames,svgnames,x11names]{xcolor} 
\usepackage{soul} 
\usepackage{subfigure}
\usepackage{float}

\usepackage{multirow,bigstrut}
\newsavebox{\measurebox}

\usepackage{hyperref}
\usepackage{natbib}
\usepackage{hyperref}
\definecolor{dark-red}{rgb}{0.9,0.15,0.15}
\definecolor{dark-blue}{rgb}{0.15,0.15,0.4}
\definecolor{dark2-blue}{rgb}{0.15,0.15,0.8}
\definecolor{medium-blue}{rgb}{0,0,0.5}
\definecolor{green}{rgb}{0,0.5,0.25}
\hypersetup{
	colorlinks, linkcolor={black},
	citecolor={dark-blue}, urlcolor={medium-blue}
}
\begin{document}

\title{Electronic states in superconducting type-II Dirac semimetal: 1T-PdSeTe}

\author{Yogendra Kumar}
\email{d225570@hiroshima-u.ac.jp}
\affiliation{Graduate School of Advanced Science and Engineering, Hiroshima University, Higashi-Hiroshima 7390046, Japan}
\affiliation{Research Institute for Synchrotron Radiation Science (HiSOR), Hiroshima University, Higashi-Hiroshima 739-0046, Japan}
	
\author{Shiv Kumar}
\email{shivam.physics@gmail.com}
\affiliation{Research Institute for Synchrotron Radiation Science (HiSOR), Hiroshima University, Higashi-Hiroshima 739-0046, Japan}
	
\author{Venkateswara Yenugonda}
\affiliation{Department of Physics, SUNY Buffalo State University, New York 14222, U.S.A}
\affiliation{Department of Physics, Indian Institute of Technology, Kanpur 208016, India}
	
\author{Ryohei Oishi}
\affiliation{Graduate School of Advanced Science and Engineering, Hiroshima University, Higashi-Hiroshima 7390046, Japan}

\author{Jayita Nayak}
\affiliation{Department of Physics, Indian Institute of Technology, Kanpur 208016, India}

\author{Chaoyu Chen}
\affiliation{Shenzhen Institute for Quantum Science and Engineering (SIQSE) and Department of Physics, Southern University of Science and Technology (SUSTech), Shenzhen 518055, China}

\author{Ravi Prakash Singh}
\affiliation{Department of Physics, Indian Institute of Science Education and Research, Bhopal 462066, India}

\author{Takahiro Onimaru}
\affiliation{Graduate School of Advanced Science and Engineering, Hiroshima University, Higashi-Hiroshima 7390046, Japan}

\author{Yasuyuki Shimura}
\affiliation{Graduate School of Advanced Science and Engineering, Hiroshima University, Higashi-Hiroshima 7390046, Japan}

\author{Shinichiro Ideta}
\affiliation{Graduate School of Advanced Science and Engineering, Hiroshima University, Higashi-Hiroshima 7390046, Japan}
\affiliation{Research Institute for Synchrotron Radiation Science (HiSOR), Hiroshima University, Higashi-Hiroshima 739-0046, Japan}

\author{Kenya Shimada}
\email{kshimada@hiroshima-u.ac.jp}
\affiliation{Graduate School of Advanced Science and Engineering, Hiroshima University, Higashi-Hiroshima 7390046, Japan}
\affiliation{Research Institute for Synchrotron Radiation Science (HiSOR), Hiroshima University, Higashi-Hiroshima 739-0046, Japan}
\affiliation{The International Institute for Sustainability with Knotted Chiral Meta Matter (WPI-SKCM$^2$), Hiroshima University, Higashi-Hiroshima 739-0046, Japan}
\affiliation{Research Institute for Semiconductor Engineering (RISE), Hiroshima University, Higashi-Hiroshima 739-8527, Japan}

	
\begin{abstract}
We have investigated the surface and bulk electronic structures of the superconducting type-II Dirac semimetal 1T-PdSeTe. The superconducting transition temperature $T_{\text{c}} = 3.2$ K was almost twice as high as $T_{\text{c}} = 1.6$ K in 1T-PdTe$_2$. Scanning transmission electron microscopy measurements showed homogeneously mixed Se and Te atoms in the chalcogen layers, consistent with the CdI$_2$-type crystal structure. Angle-resolved photoemission spectroscopy measurements and density functional theory calculations indicated the existence of the topological surface states, and the overall band structures were similar to those of 1T-PdTe$_2$. These results suggest that CdI$_2$-type lattice symmetry dictates the band dispersion, regardless of atomic disorder in the chalcogen layers. As the electronic band dispersion and the local structures were persistent upon substitution, the enhancement of $T_{\text{c}}$ is likely associated with the chemical pressure. Our results provide insight into the effects of the solid solution on the surface and bulk electronic states as well as the superconducting transition temperature.
\end{abstract}
	
\date{\today}
\maketitle
\section{Introduction}
Topological semimetals have been the focus of recent theoretical and experimental studies \cite{RevModPhys.82.3045, neupane2014observation}. Dirac semimetals, Weyl semimetals, and topological nodal-line semimetals are the three major categories characterized by the form and degeneracy of the band crossings as well as spin texture near the Fermi level ($E_{\rm F}$) \cite{yan2017topological, lv2021experimental}. 

Weyl semimetals can be realized by breaking either the spatial inversion symmetry or the time-reversal symmetry of Dirac semimetals \cite{wan2011topological}. 
They are characterized by massless chiral fermions in the bulk behaving like magnetic monopoles in momentum space ($i.e.$, Weyl fermions) and discontinuous Fermi arcs in the topological surface states. Weyl semimetals may exhibit high-charge-carrier mobility and high negative magnetoresistance under strong electric and magnetic fields, which was discussed in terms of the Adler-Bell-Jackiw anomaly (non-conservation of chiral current in the presence of gauge fields) in quantum field theory \cite{nielsen1983adler}.

There are two types of Weyl semimetals: type I with Lorentz symmetry such as TaAs, and type II without Lorentz symmetry such as WTe$_2$ and MoTe$_2$\cite{lv2021experimental}. 
The anisotropic chiral anomaly in the type-II Weyl semimetals induces the anisotropic magneto-transport properties \cite{nag2020magneto}. 

In the search for type-II Weyl semimetals, 1T-PdTe$_2$ has attracted much attention because it is a type-II Dirac semimetal and exhibits superconductivity \cite{liu2021enhanced,cook2023observation,clark2018fermiology}. 
Due to the centrosymmetric and nonsymmorphic properties of 1T-PdTe$_2$, some of the Kramers doublets degenerate at the zone boundary. 
Furthermore, 1T-PdTe$_2$ has several tilted Dirac points in the bulk  along the $\Gamma$--A direction of the bulk Brillouin zone. 
These Dirac points are protected by the strong spin-orbit interaction because the symmetry of the $R_{5,6}^+$ states with $J=3/2$ (at the $\Gamma$ point) derived from Te 5$p_x$, 5$p_y$ orbitals cannot hybridize with the $R_4^-$ state with $J=1/2$ (at the $\Gamma$ point) derived from Te 5$p_z$ orbital due to different $J$ values \cite{bahramy2018ubiquitous}. 
In addition, 1T-PdTe$_2$ hosts several topological surface states (TSSs) in the gap of the bulk bands with inverted parity \cite{clark2018fermiology, bahramy2018ubiquitous}. The TSS is robust under perturbations that conserve the topological numbers, exhibiting high carrier mobility, the quantum spin Hall effect \cite{bernevig2006quantum}, the quantum anomalous Hall effect \cite{yu2010quantized}, Majorana fermions in the superconducting state \cite{fu2010odd}, and magnetic monopoles \cite{qi2009inducing}.

As concerns superconducting property, 1T-PdTe$_2$ was reported to be a type-II superconductor with $T_{\text{c}} = 1.64$ K \cite{liu2021enhanced,cook2023observation,clark2018fermiology}. The bulk superconductivity in 1T-PdTe$_2$ can be understood from the conventional BCS theory in which Cooper pairs are formed via the electron-phonon interaction \cite{cook2023observation,yu2010quantized}. 
The density functional theory (DFT) calculations indicated significant role of a van Hove singularity derived from the saddle point near the M point in the bulk superconductivity \cite{kim2018importance,anemone2021electron}. 
On the other hand, surface superconductivity has been identified \cite{wan2011topological,nielsen1983adler,liu2021enhanced, bernevig2006quantum, ferreira2021strain,xia2024pressure} and the ac-susceptibility measurement showed an unusual surface sheath superconductivity, which is likely related to the TSS in 1T-PdTe$_2$ \cite{bernevig2006quantum,PhysRevMaterials.2.094001}.
Note that non-trivial Berry phase has been reported in the de Haas-van Alphen quantum oscillation measurements \cite{fei2017,das2018}.

To realize the Weyl semimetal starting from 1T-PdTe$_2$, one can break time-reversal symmetry by introducing magnetic elements or break the inversion symmetry by modifying the lattice \cite{xiao2018inversion}. In the case of 1T-PdTe$_2$, it has been suggested to completely substitute one of the Te layers in 1T-PdTe$_2$ with the Se layer to eliminate inversion symmetry point, and, indeed, the density functional theory (DFT) calculations indicated lifting of band degeneracy near the zone boundary\cite{bahramy2018ubiquitous}.

Recently, the superconducting transition temperature of 1T-PdSeTe was reported to be $T_{\text{c}} = 2.74$ K \cite{liu2021enhanced,liu2021new}, which was 1.67 times larger than that of 1T-PdTe$_2$ ($T_{\text{c}} = 1.64$ K) \cite{liu2021enhanced}. The magnetic susceptibility and heat capacity measurements showed the type-II superconductivity with large anisotropy and non-bulk superconductivity with a volume fraction of 20\% \cite{liu2021enhanced}. The reason for the enhanced $T_{\text{c}}$ was discussed based on chemical pressure effect and defects/disorder effect \cite{liu2021enhanced}. However, there has been no direct investigation into the electronic structure of 1T-PdSeTe so far.

In this study, we investigate the atomic arrangement, topological surface, and bulk electronic states in the 1T-PdSeTe solid solution system and explore their relationship to superconducting properties. An important advantage of solid solution systems is the tunability of the atomic composition: Se and Te belong to the same family and have the same number of valence electrons, but different atomic radii. Therefore, the lattice parameters can be tuned without changing the total valence electron numbers, which is practically the same as applying chemical pressure. If the lattice symmetry of the solid solution is preserved, the topological properties are likely to remain unchanged, as they are intrinsically tied to the symmetry of the system. This allows us to study how the lattice parameter affects the electronic parameters related to the superconducting properties as well as the topological surface states. On the other hand, since the disorder is inevitably introduced in solid solution systems, it is crucial to experimentally verify to what extent both bulk electronic states and topological surface states can be preserved.
\color{black}

We have revealed that the superconducting transition temperature enhances up to $T_{\text{c}} = 3.2$ K, which is almost twice as high as $T_{\text{c}} = 1.6$ K in 1T-PdTe$_2$. By scanning transmission electron microscopy (STEM), we have confirmed that the Se and Te atoms are homogeneously distributed in the chalcogen layers, but the lattice is compatible with the CdI$_2$-type, consistent with X-ray diffraction. 
Our super-cell calculation indicated preserved band degeneracy regardless of the loss of the centrosymmetric point due to the disorders in the chalcogen layers. 
Based on energy-band mapping in three-dimensional momentum space by angle-resolved photoemission spectroscopy (ARPES) using tunable synchrotron radiation, we confirmed the existence of TSSs and the overall band dispersions are similar to those in 1T-PdTe$_2$ \cite{bahramy2018ubiquitous}. 
Based on the local symmetry confirmed by STEM and energy-band dispersions observed by ARPES, the Se substitution can be regarded as applying a uniform chemical pressure which is likely related to the $T_{\text{c}}$ enhancement.
Our results provide insight into the effects of solid solution on the surface and bulk electronic states as well as the superconducting transition temperature.

\section{Experimental}
High-quality PdSeTe single crystals were synthesized by a two-step melting method. A stoichiometric amount of high-purity Pd wire (99.999\%), Se shots (99.999\%), and Te shots (99.999\%) were loaded inside a double quartz ampoule and sealed at a pressure of $4 \times 10^{-5} \text{ torr}$. The sealed quartz ampoule was slowly (50 $^\circ \text{C/hr}$) heated to 700 $^\circ \text{C}$, kept there for 72 hrs, and cooled to room temperature at 25 $^\circ \text{C/hr}$. This process was repeated three times with grounding the sample after each heat treatment to ensure homogeneous mixing. In the final step, the sample was pressed into pellets and sealed in a double quartz ampule at a pressure of $2 \times 10^{-5}$ Torr. The ampule was slowly heated up to 950 $^\circ \text{C}$ and kept there for 24 hrs, then slowly cooled to 550 $^\circ \text{C}$ at 3 $^\circ \text{C/hr}$, annealed for 72 hrs at 550 $^\circ \text{C}$, and then cooled to room temperature at 30 $^\circ \text{C/hr}$.

Powder and single crystal X-ray diffraction (XRD) measurements were performed on the as-grown PdSeTe sample using a Rigaku SmartLab X-ray diffractometer with Cu K$\alpha$ radiation ($\lambda = 1.54 \text{\AA}$). Powder X-ray diffraction (XRD) was carried out on both as-grown PdSeTe single crystal and powdered samples. The lattice parameters were evaluated by the Rietveld refinement of the XRD patterns using FullProf software. Laue diffraction measurements further confirmed the quality of the PdSeTe single crystal.

The elemental composition of freshly cleaved samples was studied using an electron probe micro-analyzer (EPMA) (Model no: JEOL JXA-iSP100). 

The temperature-dependent resistivity measurement was carried out using a Physical Property Measurement System (PPMS, Quantum Design Inc., USA) in the temperature range of 0.5 -- 300 K with the standard four-probe contact configuration. 

Transmission electron microscopy (TEM) and scanning TEM (STEM) measurements were done by Eurofins, EAG Laboratories. The STEM lamella with the face of $(100)$ of PdSeTe was prepared using a focused ion beam on an FEI Helios Dual Beam system and was subsequently thinned to $\sim50$ nm to achieve electron transparency. Imaging of the sample was performed using an FEI Titan S/TEM system operating at an accelerating voltage of 200 kV with $\sim 1 \text{\AA}$ resolution. Two STEM imaging modes were employed: annular dark-field STEM (ADF STEM) and bright-field STEM (BF-STEM). These modes provided complementary information on the sample's structure and composition. Additionally, energy-dispersive X-ray spectroscopy (EDS) data were acquired using Velox software, offering elemental composition information correlated with the high-resolution imaging data.

The ARPES measurement using synchrotron radiation (SR) was carried out at BL-1 at the Research Institute for Synchrotron Radiation Science (HiSOR), Hiroshima University \cite{shimada2001linear, iwasawa2017rotatable} and BL5U of UVSOR-III Synchrotron, Institute for Molecular Science \cite{ota2022uvsor}. The energy and angular resolutions were set at $\sim20$ meV and 0.1$^\circ$, respectively. The samples were cleaved {\it in situ} at $\sim20$ K using the ceramic top-post method at a base pressure better than $3.0 \times 10^{-9}$ Pa. We estimated $k_z$ assuming the nearly free-electron final-state approximation with an inner potential of $V_0 = 10$ eV:
\[
k_z = \sqrt{\frac{2m}{\hbar^2} \left(E_k \cos^2 \theta + V_0 \right)}
\]
\noindent where $E_k$ is the kinetic energy of the emitted photoelectrons.

\section{Results}
\subsection{XRD and EPMA}
\begin{figure}[htb!]
	\centering
	\includegraphics[width=8.7cm]{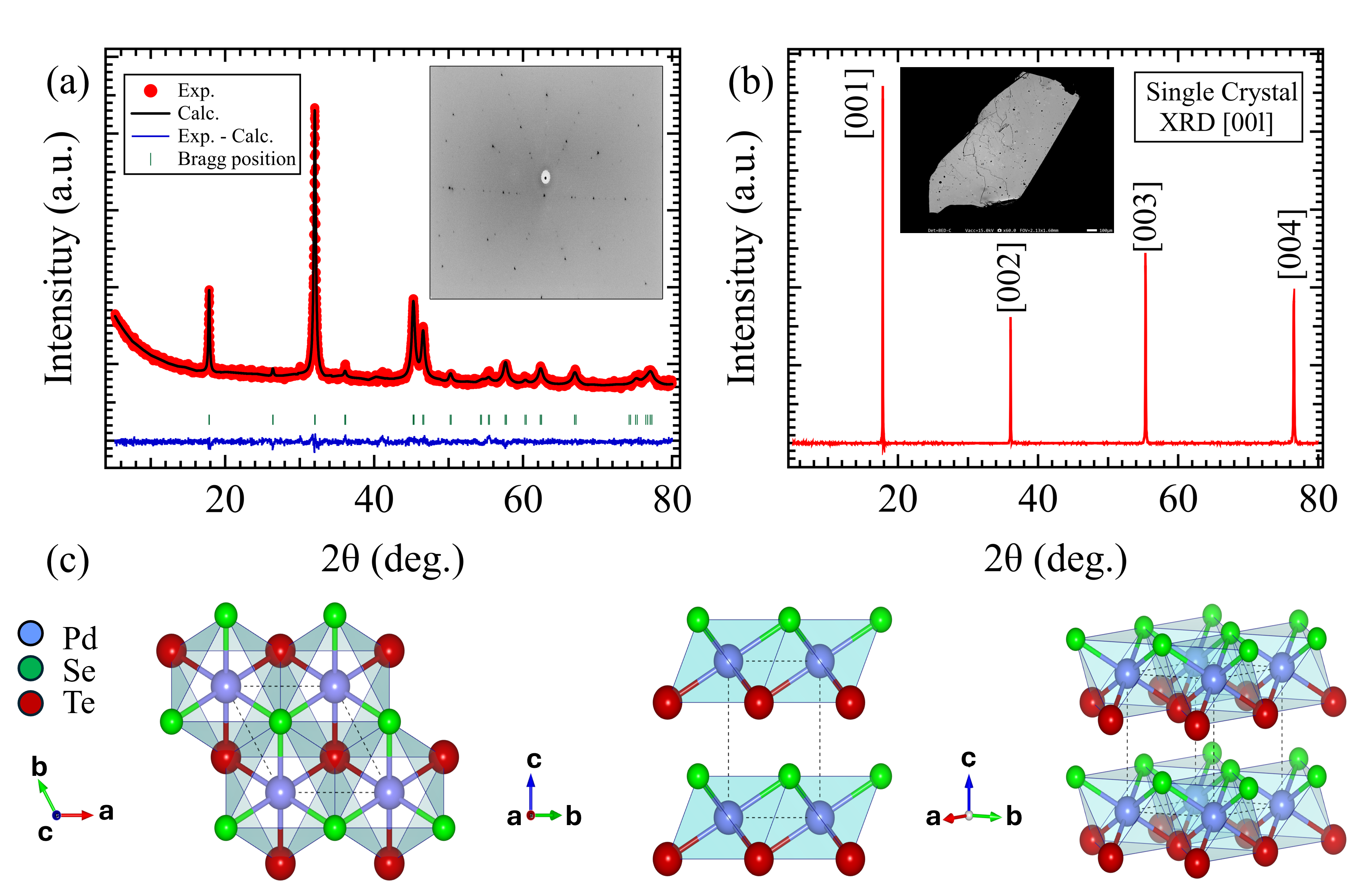}
	\caption{\label{fig:epsart}(a) Rietveld refinement profile of the XRD data for PdSeTe powder sample. The red dots represent the measured data, and the solid black line represents the calculated profile. The vertical bars in green show Bragg’s reflections for the CdI$_2$-type structure. The blue curve shows the difference between measured data and the calculated profile. Inset of (a) shows Laue’s diffraction pattern of PdSeTe on the cleaved surface. (b) Single crystal XRD data for the PdSeTe sample. The inset of (b) shows the scanning electron microscope image of the sample used for EPMA analysis. (c) Crystal structure of PdSeTe compatible with the space group P3m1 (No. 156) plotted by the VESTA software.}
	\label{fig1}
\end{figure}

Fig.~\ref{fig1} (a) and (b) respectively show the powder XRD and single crystal XRD results. 
The inset of Fig.~\ref{fig1} (a) shows Laue diffraction patterns obtained from a (001) cleaving plane ($c-$plane) exhibiting hexagonal symmetry. In the previous studies, the CdI$_2$-type crystal structure, namely the space group of P$\overline{3}$m1 (No. 164), was assumed to analyze the XRD data \cite{liu2021new, mccarron1976high}. 
If one of the Te layers in the unit cell is completely replaced by a Se layer, the inversion symmetry point will be lost, and the crystal structure should be the space group P3m1 (No. 156), as shown in Fig.~\ref{fig1} (c).
The refined $\chi^2$ values for P$\overline{3}$m1 and P3m1 are, however,  almost comparable, and we could not distinguish these two space groups. Below, we will discuss the crystal structure based on our STEM measurements.

The obtained lattice parameters were $a = b = 3.901 \text{\AA}$ and $c = 4.978 \text{\AA}$, and the cell volume was $65.605 \text{\AA}$. The calculated $\alpha$, $\beta$, and $\gamma$ angles were $90^\circ$, $90^\circ$, and $120^\circ$, which were consistent with previous studies \cite{liu2021enhanced, hulliger1965electrical}. As the atomic radius of Se is smaller than that of Te, the lattice parameters $a$ and $c$ of PdSeTe are smaller than those of 1T-PdTe$_2$ by 3.49$\%$ and 3.47$\%$, respectively. However, the value of $c/a \approx 1.27$ was the same for PdSeTe and 1T-PdTe$_2$. Since the $c/a$ ratio is less than the ideal hcp lattice value $\sqrt{8/3} \approx 1.63$ \cite{finlayson1986lattice, noh2017experimental}, the local $O_h$ symmetry centered at the Pd site is lowered to the $D_{3d}$ symmetry. Due to the enhanced overlap integral along the {\it c}-axis ($\langle100\rangle$ direction), the bonding and antibonding energy splitting is assumed to be larger along the {\it c}-axis direction. Due to the decrease of lattice parameters, the overall unit cell volume of PdSeTe is reduced by 10\% compared with that of 1T-PdTe$_2$.

The EPMA measurements were done on three single crystal samples (cut from different positions of the ingot) and we obtained compositions from thirteen different positions on each sample. On average, we obtained Pd $\approx 33.56\%$, Se $\approx 31.48\%$, and Te $\approx 34.96\%$, and then the composition of the sample can be written as PdSe$_{1-\delta}$Te$_{1+\delta}$ ($\delta \approx 0.05$).

\subsection{TEM/STEM}
\begin{figure}[htb!]
	\centering
	\includegraphics[width=8.7cm] {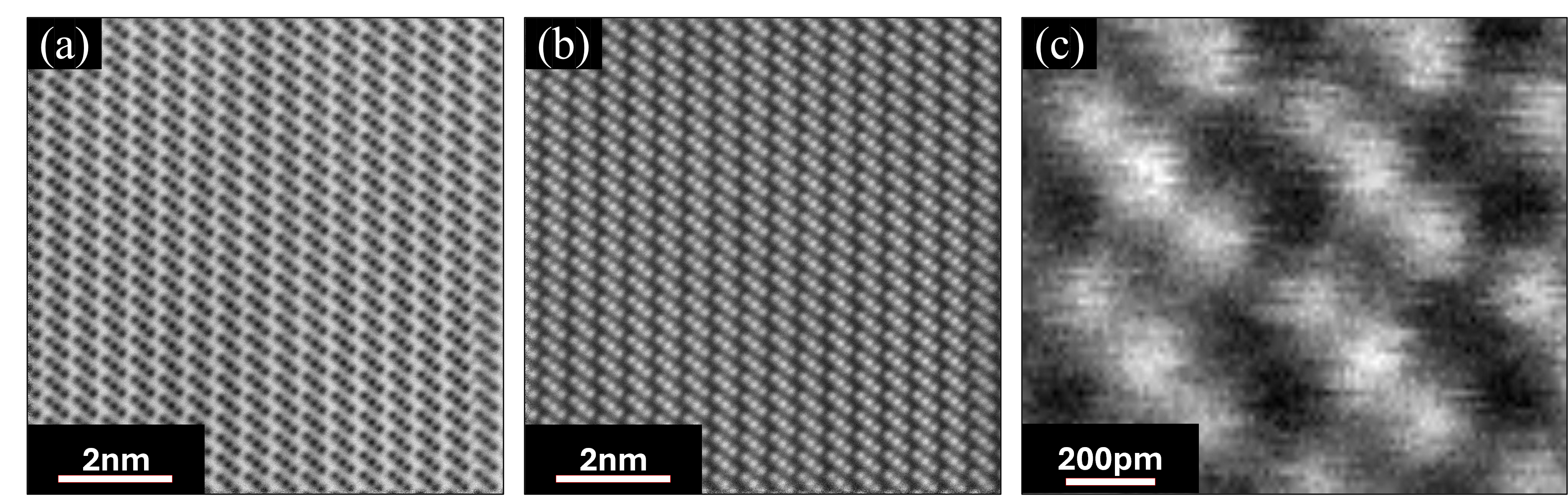}
	\caption{High-resolution STEM image of the (100) plane of a 1T-PdSeTe crystal. (a) and (b) show the bright-field image and high-angle annular dark-field (HAADF) image, respectively. (c) Magnified atomic resolution HAADF image.}
	\label{fig2}
\end{figure}

To visualize the real space crystal structure and identify atom species, we have measured STEM imaging with atomic resolution for the (100) plane of PdSeTe. Fig. \ref{fig2} shows a high-angle annular dark-field scanning transmission electron microscopy (HAADF-STEM) image of PdSeTe for the (100) cross-section. Here, the incident electron beam is parallel to the $a$-axis, namely, $\langle100\rangle$ direction. As the sample thickness is $\sim 50$ nm $\approx 123a$, the electron transmission intensity is averaged over $\approx 123$ atoms along the $\langle100\rangle$ direction. Within about $10 \times 10 \text{ nm}^2$ area, we observed a highly homogeneous lattice image consistent with the CdI$_2$ structure. 
Based on the length of 15--22 unit cells in the TEM image in Fig. \ref{fig2} (b), we have estimated $a=4.05 \text{\AA}^{-1}$ and $c=5.12 \text{\AA}^{-1}$, which were only 3--4$\%$ larger than the XRD data.  This level of agreement is remarkable, considering the instrumental resolution of $\sim 1\text{\AA}$ for STEM.

\begin{figure}[htb!]
	\centering
	\includegraphics [width=8.7cm] {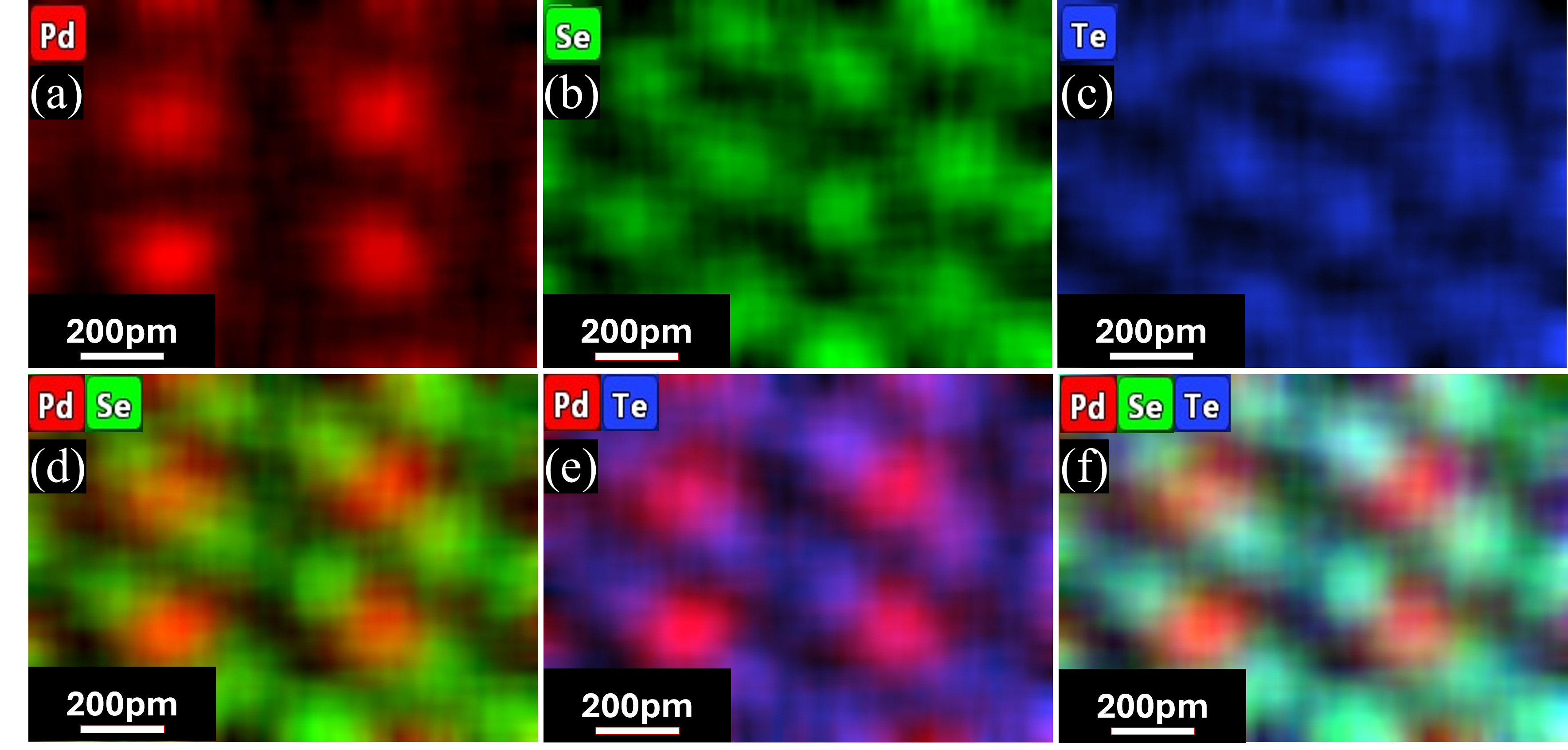}
	\caption{Atomically resolved HAADF-STEM EDX mappings of PdSeTe single crystal. (a), (b), and (c) show signals from Pd (red), Se (green), and Te (blue) atoms, respectively. (d) shows the overlaid image of (a) and (b). (e) shows the overlaid image of (a) and (c). (f) shows the overlaid image of (a), (b), and (c).}
	\label{fig3}
\end{figure}

In Figs. \ref{fig3}(a)--\ref{fig3}(c), the energy-dispersive X-ray spectroscopy (EDX) maps show the elemental distribution of Pd (red), Se (green), and Te (blue) obtained by STEM. 
The Pd and Se contributions are overlaid in Fig. \ref{fig3}(d) and Pd and Te contributions are overlaid in Fig. \ref{fig3}(e). It is clearly seen that Pd atoms are well separated from the Se and Te atoms, indicating that Pd layers are occupied by Pd atoms only. On the other hand, Figs. \ref{fig3}(b) and \ref{fig3}(c) indicate Se and Te atoms almost overlap, suggesting these atoms equally occupy the chalcogen layers. In Fig. \ref{fig3}(f), Pd, Se, and Te contributions are all overlaid, and one can confirm Se and Te atoms are mixed in the chalcogen layers. To further confirm the results, we checked the intensity distribution of the HAADF-STEM image which is sensitive to atomic number ($Z$). As Te is heavier than Se, the intensity should be stronger. However, the intensity of chalcogen layers is highly homogeneous indicating that Se and Te atoms are evenly mixed in the chalcogen layers.

\subsection{Resistivity and Magnetic Susceptibility}

\begin{figure}[htb!]
	\centering
	\includegraphics [width=8.7cm] {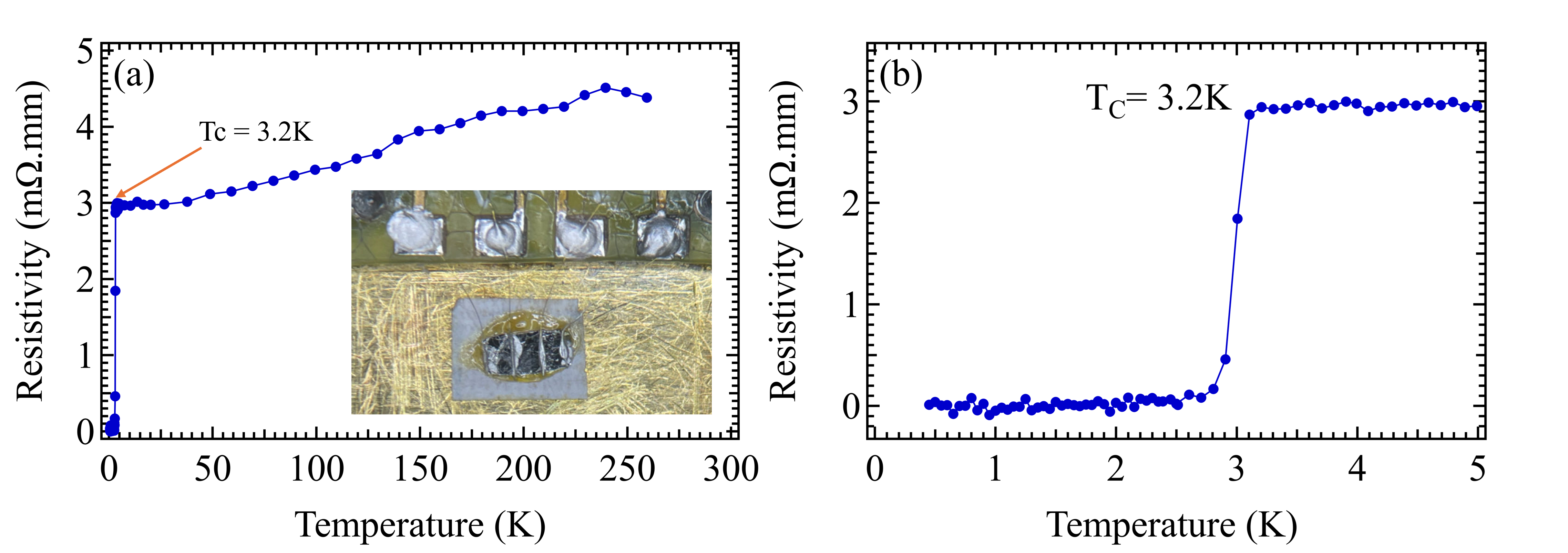}
	\caption{(a) Resistivity as a function of temperature (Inset: four probe connections on PPMS sample stage). (b) Enlarged view of resistivity near the superconducting transition temperature.}
	\label{fig4}
\end{figure}

Figure \ref{fig4}(a) shows the temperature dependence of the electrical resistivity ($\rho$) from 0.5 K to 300 K. In Fig. \ref{fig4}(b), the resistivity abruptly drops to zero at the superconducting transition temperature ($T_{\text{c}}$) of 3.2 K. The observed $T_{\text{c}}$ is almost twice as high as that of PdTe$_2$ (1.64 K) and even larger than the value, 2.74 K, previously reported for PdSeTe \cite{liu2021enhanced}.

\begin{figure}[htb!]
	\centering
	\includegraphics [width=8.7cm]{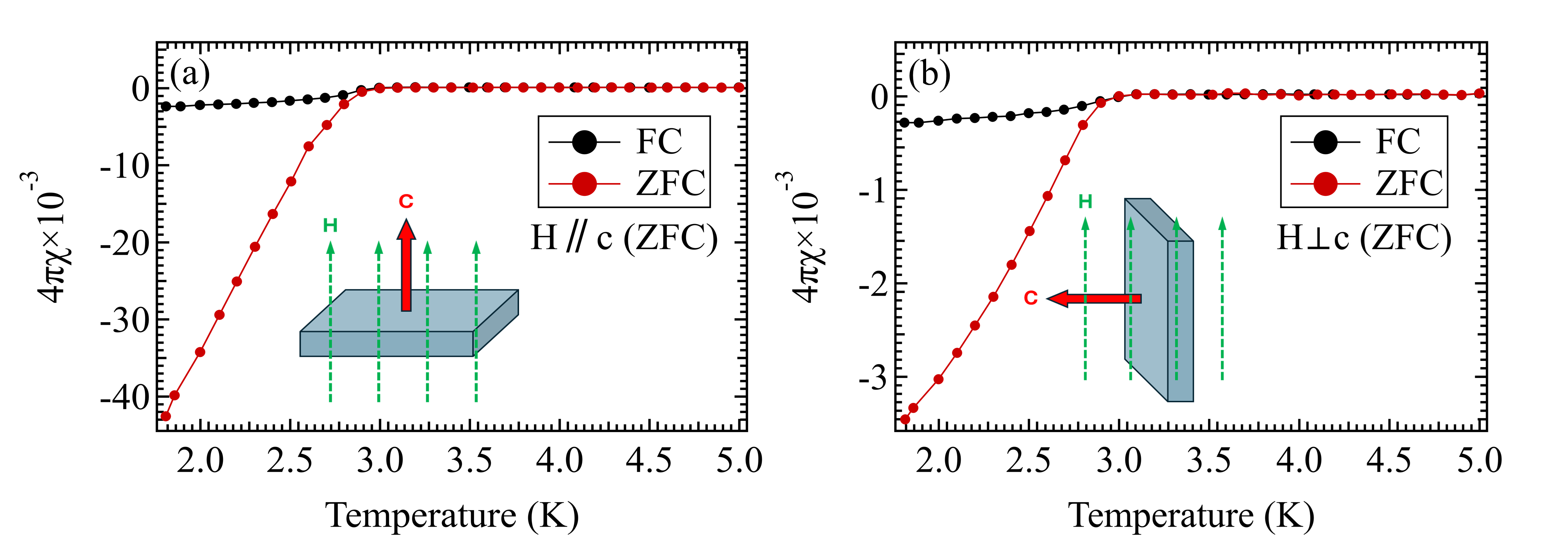}
	\caption{Temperature dependence of magnetic susceptibility measured in FC and ZFC mode with an external field of 1.5 mT. (a) Magnetic field is parallel to the $c$-axis ($H \parallel c$). (b) Magnetic field is perpendicular to the c axis ($H \perp c$).}
	\label{fig5}
\end{figure}

Figs. \ref{fig5}(a) and \ref{fig5}(b) present the temperature-dependent magnetic susceptibility measured in zero-field-cooled (ZFC) and field-cooled (FC) modes under an external magnetic field ($H$) of 1.5 mT, applied parallel to the $c$-axis and perpendicular to the $c$-axis, respectively. In both modes, a superconducting transition was observed at 3.2 K, consistent with the resistivity data. The superconducting volume fraction was estimated to be 95.5\% for $H \parallel c$ [Fig. \ref{fig5}(a)] and 90.0\% for $H \perp c$ [Fig. \ref{fig5}(b)]. We assume that the $\sim 5\%$ difference in the susceptibility is due to the possible difference in mobility and pinning of the vortices within the layered crystal structure. For $H \parallel c$, the vortices can effectively shield the magnetic field due to the stronger in-plane conduction, whereas for $H \perp c$, the weaker interlayer coupling makes the shielding less effective.

\subsection{Electronic structure calculation}
To understand the electronic structure of PdSeTe, we have performed density functional theory (DFT) calculations. 
We used the FLEUR DFT package \cite{wortmann2023fleur}, which utilizes a full-potential linearized augmented plane-wave and local orbitals (FP-LAPW+lo) basis \cite{PhysRevB.43.6388,wimmer1981full,betzinger2010hybrid} 
along with the Perdew-Burke-Ernzerhof parametrization of the generalized gradient approximation as the exchange-correlation functional \cite{betzinger2010hybrid}. 
The plane-wave cutoff parameter was set to five times the largest muffin-tin radius, and the irreducible Brillouin zone was sampled by a $10 \times 10 \times 12$ mesh of $k$-points for the bulk calculations and an $18 \times 18 \times 1$ mesh of $k$-points for the slab model calculations.

The lattice parameters used in the calculations, $a = b = 3.901 \text{\AA}$, and $c=4.978 \text{\AA}$ were taken from our Rietveld refinement analyses of the powder X-ray diffraction patterns. 
\color{black}
Spin-orbit coupling was included as a second variational procedure. Irreducible representations were determined using the built-in irrep tool in FLEUR \cite{wortmann2023fleur}.

\begin{figure}[htb!]
	\centering
	\includegraphics [width=8.7cm]{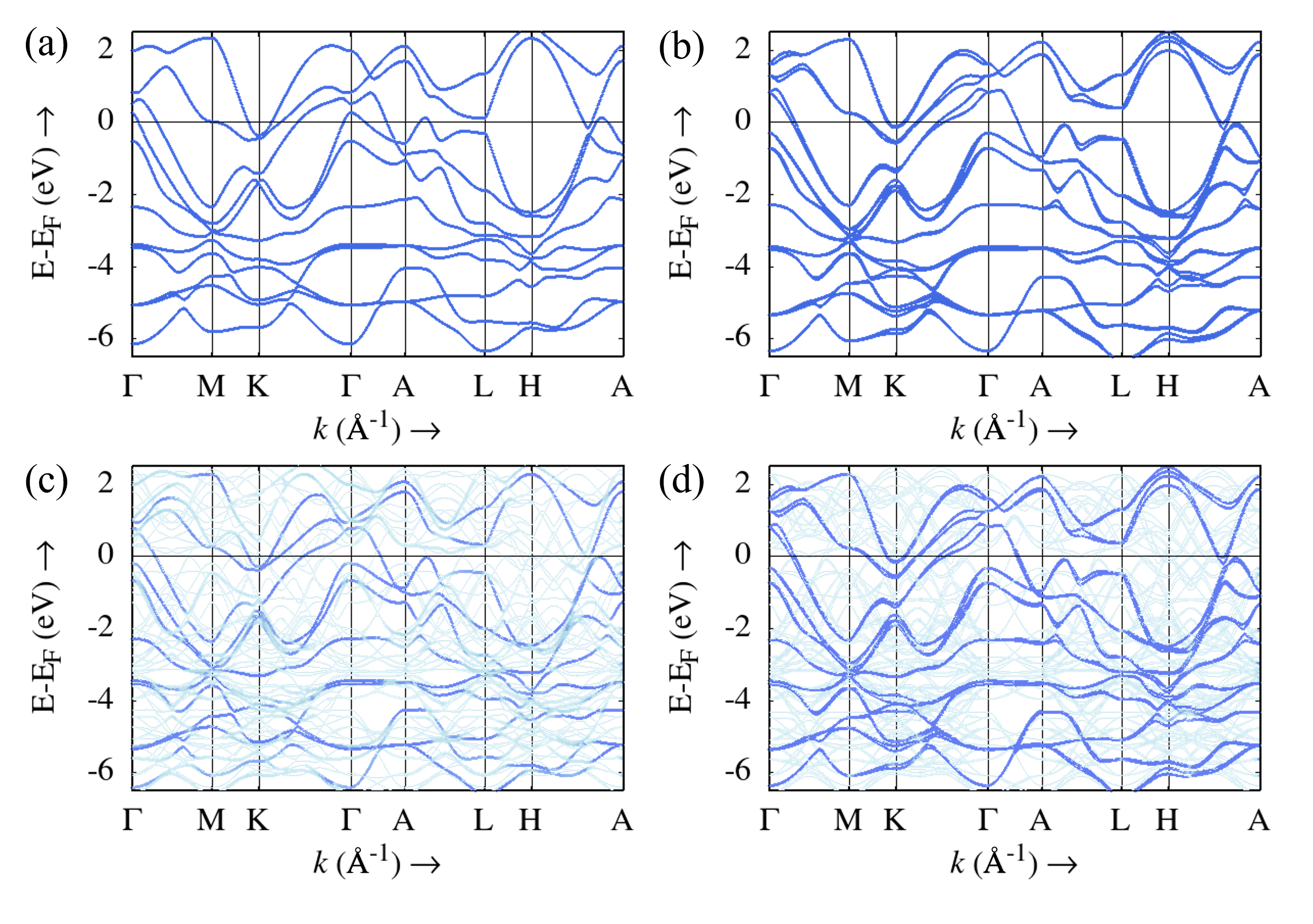}
	\caption{(a) The bulk DFT calculations of PdTe$_2$ with the space group of P$\overline{3}$m1 (No. 164). (b) The bulk DFT calculations of PdSeTe with the space group of P3m1 (No. 156). (c) The DFT calculations of the bulk bands on $2 \times 2 \times 2$ super-cell of PdSeTe where an even number of the Se and Te atoms occupy the chalcogen layer, breaking the space group symmetry of P3m1 (No. 156). (d) The DFT calculations on a $2 \times 2 \times 2$ super-cell but with the well-defined Se and Te layers to hold the space group symmetry of P3m1 (No. 156). The energy bands in (c) and (d) are unfolded into the Brillouin zone of the CdI$_2$ type crystal structure. The unfolded energy bands in (d) are consistent with those of (b).}
	\label{fig6}
\end{figure}

Fig. \ref{fig6}(a) and Fig. \ref{fig6}(b) respectively show the DFT calculation results of the bulk bands of 1T-PdTe$_2$ with the space group of P$\overline{3}$m1 (No. 164) and those of PdSeTe with the space group of P3m1 (No. 156). 

From Figs. 6(a) and 6(b), one can see some significant changes in the bulk band structures near the $E_\text{F}$. The saddle point around the M point and the electron-like band around the K point in PdSeTe [Fig. 6(b)] shift upward compared to those in PdTe$_2$ [Fig. 6(a)]. Around the $\Gamma$ point, the number of hole-like bands crossing the $E_\text{F}$ decreases from PdTe$_2$ [Fig. 6(a)] to PdSeTe [Fig. 6(b)]. Along the A--H and A--L lines, on the other hand, small hole-like bands do not cross the $E_{\rm F}$ in disordered PdSeTe.
\color{black}
Due to the loss of the centrosymmetric point in PdSeTe, the degeneracy of the Kramers doublet protected by the inversion symmetry is lifted along most of the high symmetry directions except for the $\Gamma$--A direction. 
{For example, in PdSeTe [Fig. 6(b)], the unoccupied bands along the $\Gamma$--K direction show band splittings that are absent in PdTe$_2$ [Fig. 6(a)].}
We should note that along the $\Gamma$--A direction, the basis function is 1 or $z$ and the $C_{3v}$ symmetry properties are not affected by the loss of the inversion symmetry point.

To understand the disorder effect in the chalcogen layers as revealed by STEM measurements, we have calculated the band structure for a $2 \times 2 \times 2$ super-cell 

where the same number of Se and Te atoms occupy the chalcogen layer, 
breaking the symmetry of P3m1 (No. 156) in Fig. \ref{fig6}(c). 
As a reference, we also performed the DFT calculation on the $2 \times 2 \times 2$ super-cell that holds the symmetry of the space group P3m1 (No. 156) in Fig. \ref{fig6}(d), which should give results consistent with the standard bulk calculation as shown in Fig. 6(b).
\color{black}
The energy bands in Figs. \ref{fig6}(c) and \ref{fig6}(d) are unfolded into the Brillouin zone of the CdI$_2$-type crystal structure.

One can confirm that the band dispersions shown in Figs. \ref{fig6}(b) and \ref{fig6}(d) are identical for PdSeTe
{ because the symmetry properties [P3m1 (No. 156)] are preserved.}
On the other hand, in Fig. \ref{fig6}(c), the super-cell calculation with disordered chalcogen layers does not show splitting 

as seen in the unoccupied band along the $\Gamma$--K direction in Figs. \ref{fig6}(b) and \ref{fig6}(d).
In terms of the absence of such band splitting, the situation is similar to that of the band dispersion of PdTe$_2$. 
This suggests that the inversion symmetry is effectively restored when the Se and Te atoms are randomly arranged in the chalcogen layer by 50\%.
By detailed comparison between Figs. \ref{fig6}(a) (PdTe$_2$) and \ref{fig6}(c) (disordered PdSeTe) near the $E_{\rm F}$, in disordered PdSeTe, one of the hole-like bands around the $\Gamma$ point is completely occupied, and the saddle point around the M point is elevated. Along the A--H and A--L lines, on the other hand, hole-like bands do not cross the $E_{\rm F}$ in disordered PdSeTe. 
We assume that these results effectively demonstrate the band modifications caused by compressing the lattice parameter while maintaining lattice symmetry, namely, the chemical pressure effect.
\color{black}

\begin{figure}[htb!]
	\centering
	\includegraphics [width=8.7cm] {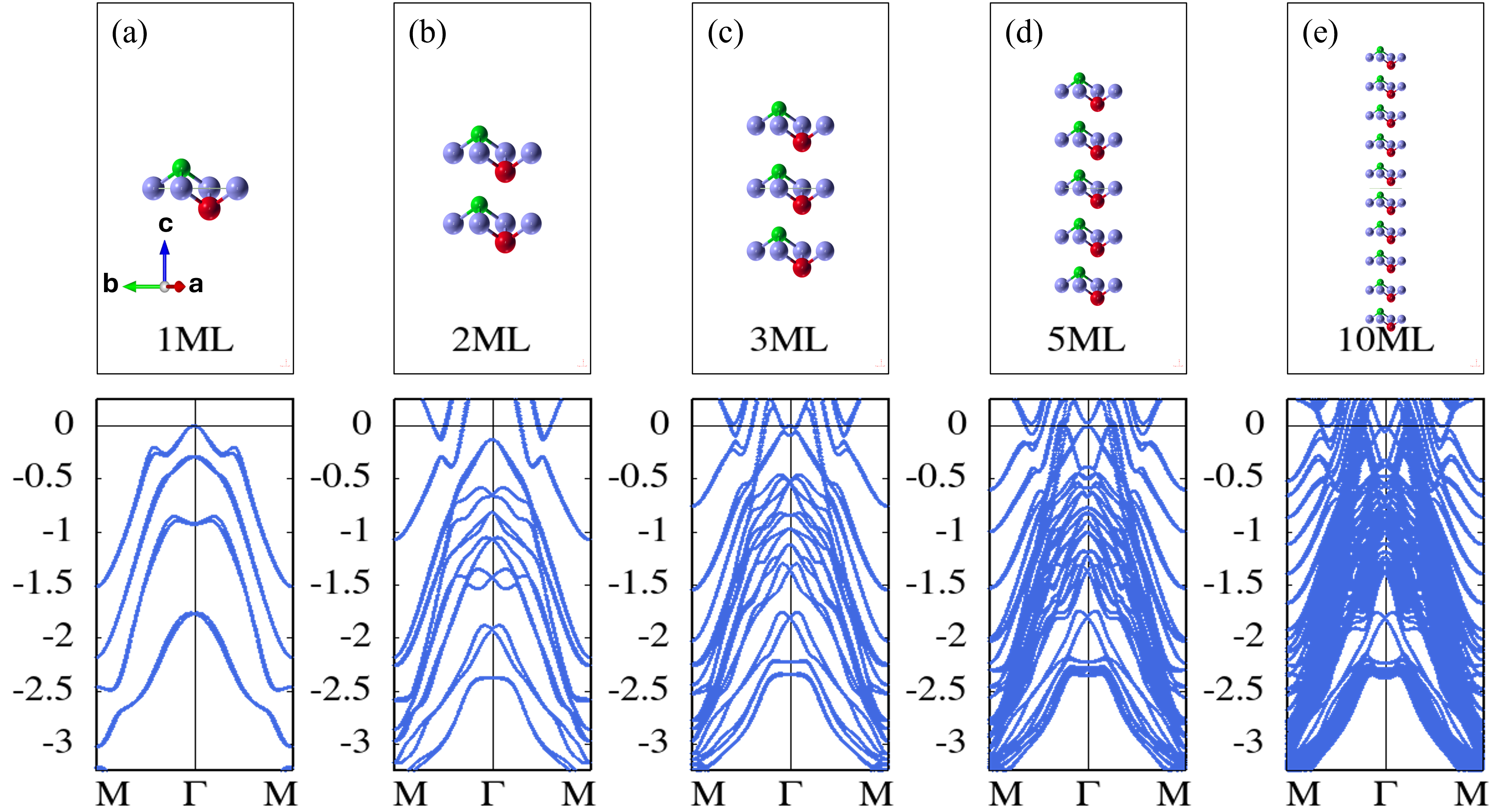}
	\caption{Thickness-dependent DFT calculations of PdSeTe with (a) 1ML, (b) 2ML, (c) 3ML, (d) 5ML and (e) 10ML.}
	\label{fig7}
\end{figure}

To understand the surface-derived states, we have done DFT calculations on slab models for different monolayer (ML). Here we did not consider the disorder effect in the chalcogen layers. We assume the disorder effect does not significantly alter the band positions based on the results of super-cell calculations in Figs. \ref{fig6}(c) and \ref{fig6}(d). As shown below the results gave reasonable explanations for the ARPES results.
The slab was constructed by stacking five unit-cells along the $c$-direction with a $4.978 \text{\AA}$ vacuum gap. 

Figs. \ref{fig7}(a)-\ref{fig7}(e) show the calculated results from 1ML up to 10ML. 
At the $\overline{\Gamma}$ point and at the energy of $-1.5$ eV and $-1.8$ eV, one can clearly see Dirac-fermion-like crossings which should correspond to the topological surface states (TSS1 and TSS2 reported in PdTe$_2$ \cite{cook2023observation,clark2018fermiology, bahramy2018ubiquitous}). The former TSS falls in the bulk band projection but the latter TSS resides within the gap of the bulk band projection. 
Due to the parity inversion in the bulk bands, these two states are topologically non-trivial \cite{fu2010odd}. 
In our calculation, the Dirac point at $-1.8$ eV is robust even in 1ML while recent thickness-dependent DFT calculations on PdTe$_2$ indicated gap opening at the Dirac point below 3ML \cite{cook2023observation}. 
Note that the electron-like topologically nontrivial surface states exist near the $E_{\rm F}$ around the middle points of the $\overline{\Gamma}$--$\overline{\rm M}$ line.

\subsection{Angle-resolved photoemission spectroscopy}
 
\begin{figure}[htb!]
	\centering
	\includegraphics [width=8.7cm]{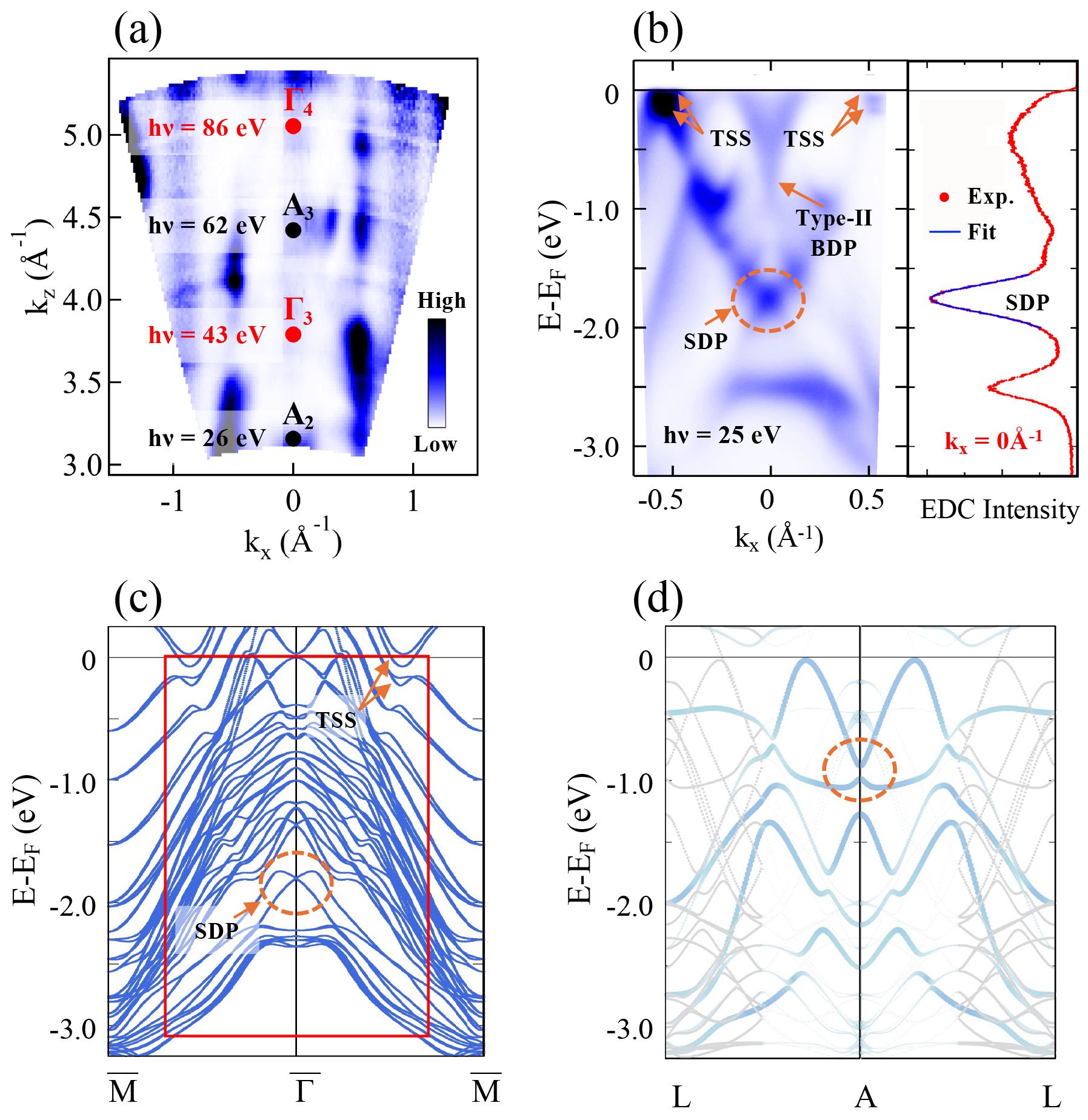}
	\caption{(a) Fermi surface cross section of PdSeTe by the $ k_x- k_z$ plane.  Here $k_x$ and $k_z$ directions are parallel to $\overline{\Gamma}$--$\overline{\rm M}$ direction and $\Gamma$--A direction, respectively. High symmetries points (A$_2$, $\Gamma_3$, A$_3$, and $\Gamma_4$) and required photon energies to reach these points are shown. (b) ARPES intensity plot near the L--A--L direction (left panel) and energy distribution curve (EDC) along $k_x = 0\text{\AA}^{-1}$ (right panel) taken at $h\nu = 25$ eV. The dashed circle shows the surface-derived Dirac point (SDP), and arrows show the topological surface states (TSS) and bulk-derived type-II Dirac point (BDP). The blue line in the right panel shows a Lorentzian fit. (c) The DFT calculation result for 5 ML along the $\overline{\rm M}$--$\overline{\Gamma}$--$\overline{\rm M}$ high-symmetry direction. The red box corresponds to the range of the ARPES image in (b). The SDP and TSS corresponding to those in (b) are also shown. (d) The DFT calculation result of the bulk band dispersion along L--A--L direction obtained from a super-cell with disordered Se and Te atoms. The degeneracy of the BDP is lifted just at the A point (see dashed circle part). However, the BDP exists along the $\Gamma$-A direction, closer to the A point.}
	\label{fig8}
\end{figure}

We investigated the electronic band structure of PdSeTe via ARPES measurements. Fig. \ref{fig8}(a) shows the Fermi surface cross-section in the $k_x - k_z$ plane, obtained using various photon energies ranging from $h\nu=25$ eV to 100 eV. 
Here, the $k_x$ and $k_z$ directions are parallel to the $\overline{\Gamma}$--$\overline{\rm M}$ direction and the $\Gamma$--A direction, respectively. 
The $k_z$ region contains high-symmetry points along the A--$\Gamma$--A direction in the normal emission geometry, from the second A$_2$ point ($h\nu=26$ eV) up to the fourth $\Gamma_4$ point ($h\nu=86$ eV) in the extended zone scheme [Fig. 8(a)].
We determined the inner potential based on the weak but finite $k_z$ periodicity of several bands. 
Fig. \ref{fig8}(b) shows the ARPES intensity map taken at $h\nu=25$ eV, which corresponds near the A$_2$ point at $k_x=0$ \AA$^{-1}$. Thus, the $k_x$ direction is close to along the L--A--L direction. 
In Fig. 8(c), one can see surface-derived Dirac point  at $-1.75$ eV below the $E_{\rm F}$ [inside the dashed circle], type-II bulk-derived Dirac point at $-0.93$ eV, and two split topological surface state just below the Fermi level around $k_x = \pm 0.5 \text{\AA}^{-1}$. 
These spectral features all exist in the ARPES results of PdTe$_2$\cite{cook2023observation,clark2018fermiology,bahramy2018ubiquitous}.
In the right panel of Fig. 8(b), we extracted an energy distribution curve and fitted the peak of the surface-derived Dirac point using a Lorentzian [blue line in the figure]. 
As no significant peak splitting was detected, we assume the surface state to be topologically non-trivial.
The observed topological surface state just below the $E_\text{F}$ and the surface-derived Dirac point are consistent with our 5ML slab calculation in Fig. \ref{fig8}(c). 

While the spectral intensity of the topological surface states are stronger than that of the bulk states, the location of the type-II Dirac point ($-0.93$ eV) in the bulk is consistent with that of PdTe$_2$ measured near the A point taken at $h\nu=60$ eV \cite{cook2023observation,clark2018fermiology,bahramy2018ubiquitous}. 

\color{black}
Fig. \ref{fig8}(d) shows the theoretical band dispersion along the L--A--L high symmetry line for a super-cell with disordered Se and Te atoms in the chalcogen layer. 
The electron-like band near the $E_{\rm F}$ centered at the A point in Fig. \ref{fig8}(d) is consistent with the observed electron-like band in Fig. \ref{fig8}(b). 

It should be noted that the bulk Dirac point is slightly away from the A point, so the degeneracy was lifted just at the A point [see the dashed circle region in Fig. 8(d)]. 
\color{black}
The flat band at the $\overline{\Gamma}$ point at $-2.5$ eV in Fig. \ref{fig8}(b) corresponds to that in Fig. \ref{fig8}(c). Fig. \ref{fig8}(d) also exhibits corresponding spectral features around $-2.5$ eV, but with somewhat larger dispersion around the A point.

\begin{figure}[htb!]
	\centering
	\includegraphics [width=8.7cm] {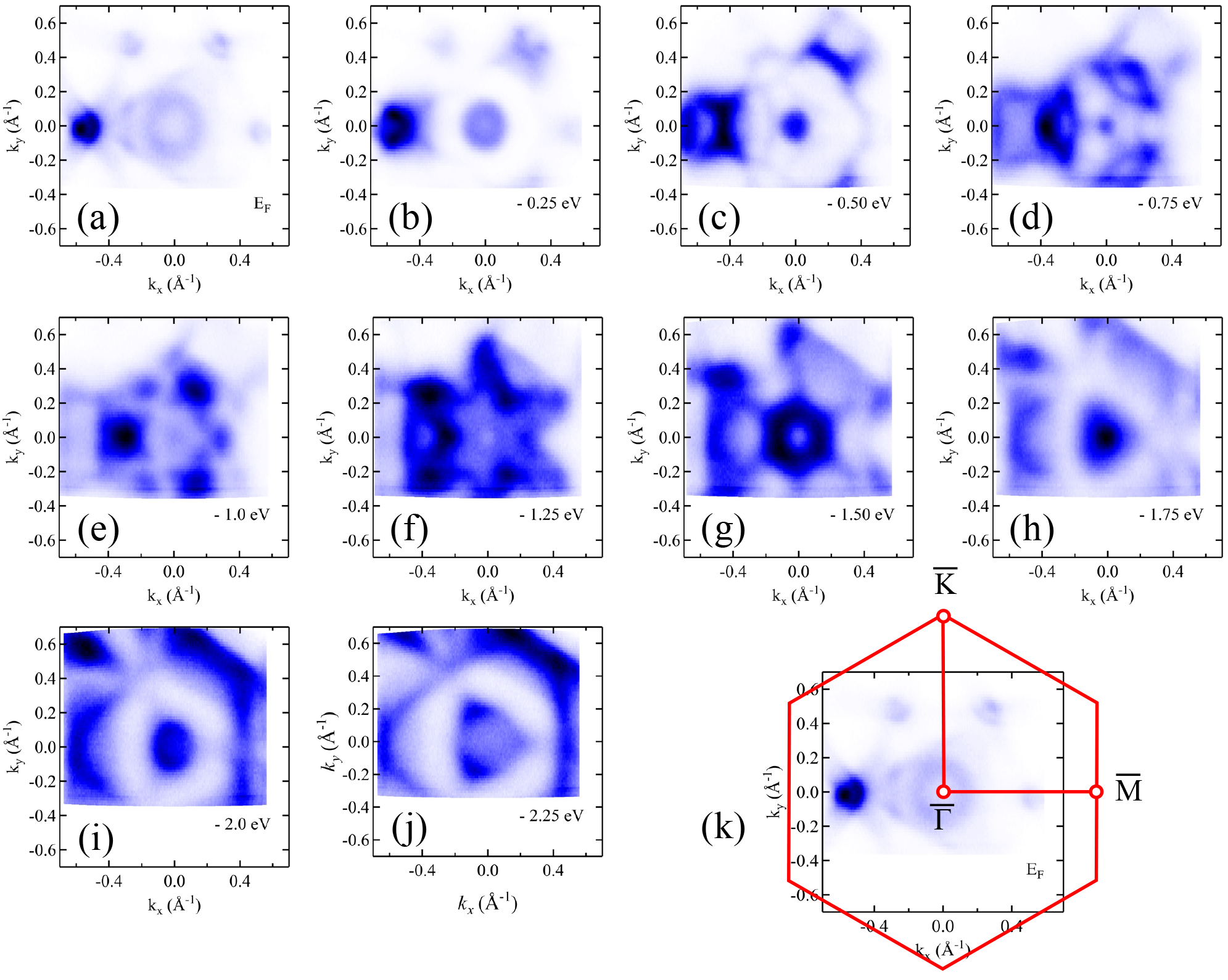}
	\caption{(a)--(j) Constant energy contour from the $E_{\rm F}$ down to $-2.25$ eV measured at { $h\nu=25$ eV}. Here $k_x$ and $k_y$ directions are parallel to $\overline{\Gamma}$--$\overline{\rm M}$ direction and $\overline{\Gamma}$--$\overline{\rm K}$ direction, respectively. The bottom of the electron-like band centered at the $\overline{\Gamma}$ point exists at the energy of $-0.75$ eV. The Dirac point of the topological surface state exists at the $\overline{\Gamma}$ at the energy of $-1.75$ eV. (k) The first Brilluoin zone overlaid on (a).}
	\label{fig9}
\end{figure}

Figs. \ref{fig9}(a)--\ref{fig9}(j) show constant energy contours from the $E_{\rm F}$ down to $-2.25$ eV. 
Here, the $k_x$ and $k_y$ directions are parallel to the $\overline{\Gamma}$--$\overline{\rm M}$ direction and the $\overline{\Gamma}$--$\overline{\rm K}$ direction, respectively. 
While the Fermi surface intensity distribution in Fig. \ref{fig9}(a) appears nearly six-fold symmetric, there exists three-fold symmetric intensity modulation for lower energies, which is likely due to the matrix element effect.

Taken at $h\nu=25$ eV, the $\overline{\Gamma}$ point of the surface Brillouin zone is close to the A point of the bulk Brillouin zone. The circular Fermi surface around the $\overline{\Gamma}$ point matches the bulk derived electron-like band as shown in Fig. \ref{fig8}(d). With lowering energy, this Fermi surface shrinks to a point at an energy of $-0.75$ eV, because the bottom of the electron-like energy band exists at $-0.93$ eV as shown in Fig. \ref{fig8}(b). 
\color{black}
On the other hand, in Fig. \ref{fig9}(h), one can see the Dirac point of the topological surface state at the $\overline{\Gamma}$ point at the energy of $-1.75$ eV as shown in Fig. \ref{fig8}(b).

\begin{figure}[htb!]
	\centering
	\includegraphics [width=8.7cm]{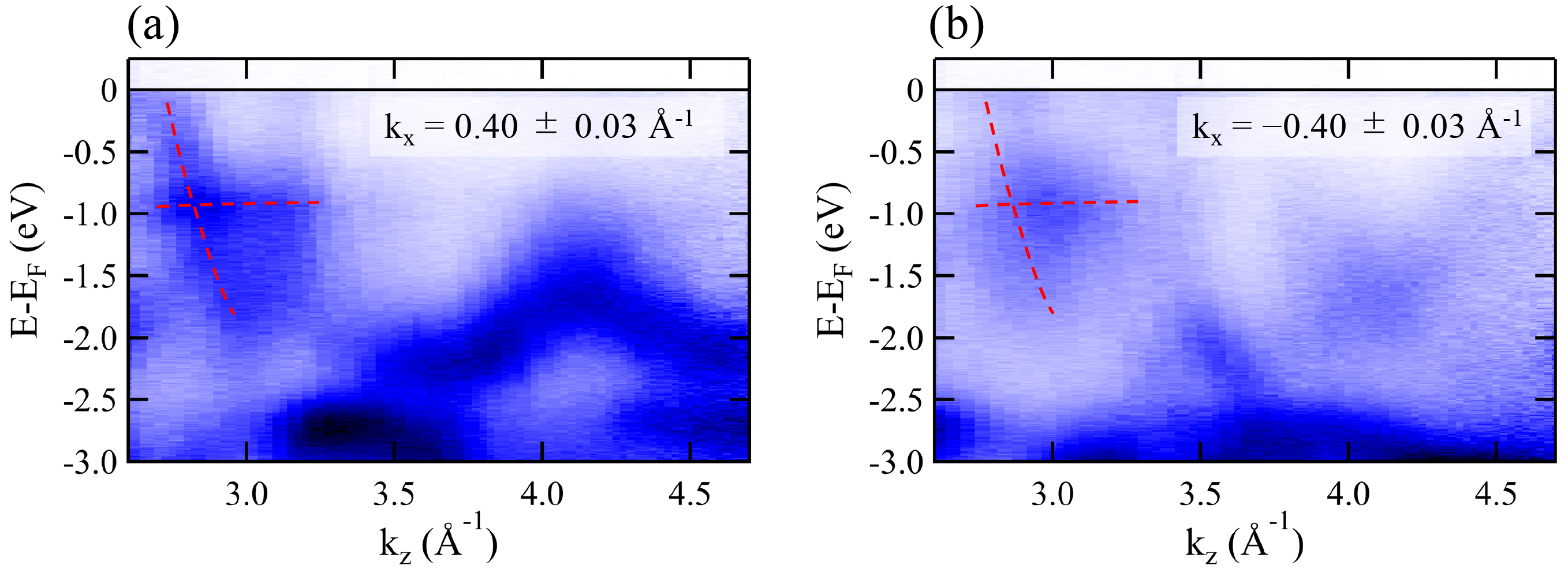}
	\caption{Band dispersion along the A--$\Gamma$--A direction with (a)$k_x = 0.40 \pm 0.02 \text{\AA}^{-1}$ and (b)$k_x = -0.40 \pm 0.02 \text{\AA}^{-1}$.  Dashed lines are guides for the eyes.}
	\label{fig10}
\end{figure}

To obtain further evidence of the bulk-derived band, we show band dispersion along $k_z$ direction at $k_x = -0.40 \text{\AA}^{-1}$ in Fig. \ref{fig10}(a) and at $k_x = 0.40 \text{\AA}^{-1}$ in Fig. \ref{fig10}(b). While the intensity is much stronger in Fig. \ref{fig10}(a) than that in Fig. \ref{fig10}(b) which is likely due to the matrix element effect, we assume the dispersive features existing between $-1.0$ eV and $-2.0$ eV should be the same. We also detect $k_z$-independent dispersion at $-1.0$ eV around $k_z = 2.8  \text{\AA}^{-1}$, exhibiting a band crossing similar to the tilted Dirac point.

\begin{figure}[htb!]
	\centering
	\includegraphics [width=8.7cm]{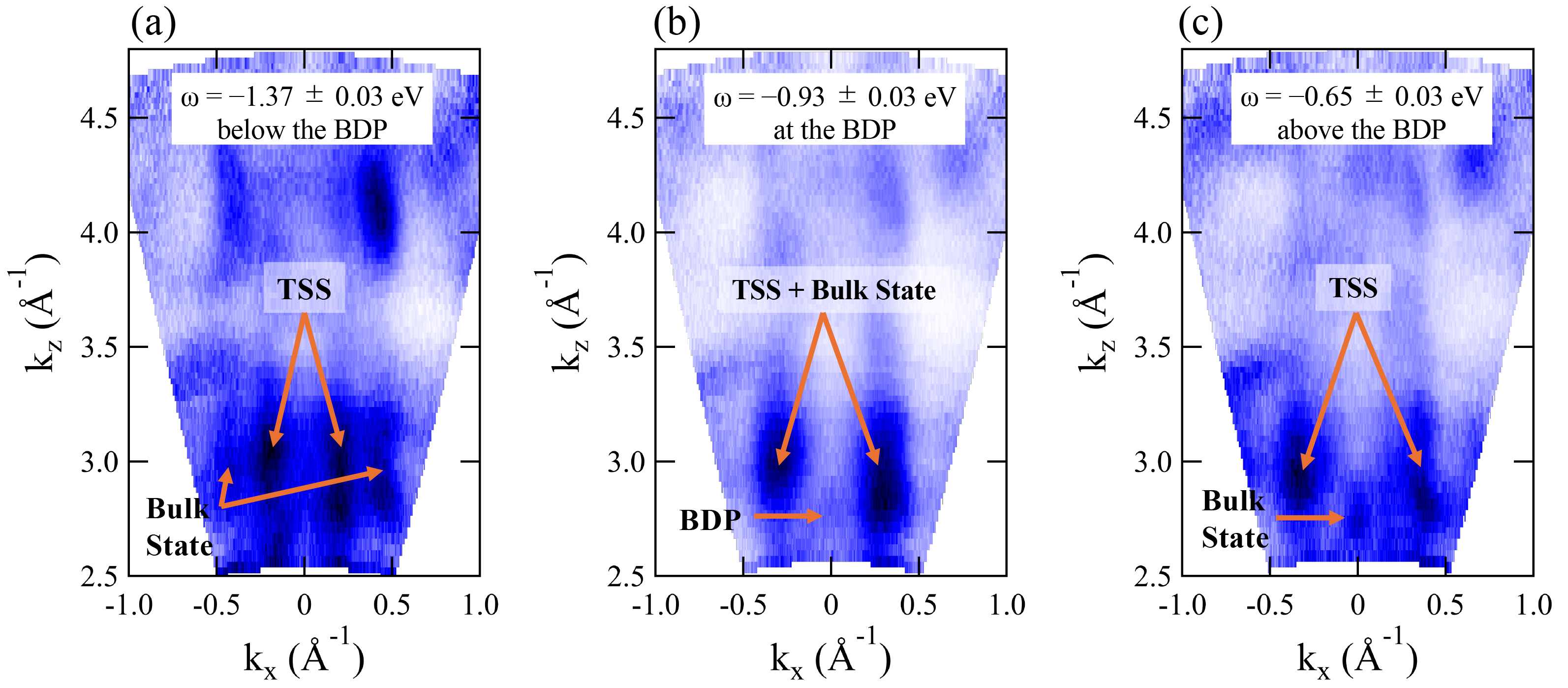}
	\caption{$k_x - k_z$ dispersion along $\Gamma$--A direction (a) band dispersion below the Dirac point energy (b) at the Dirac point energy (c) above the Dirac point energy.}
	\label{fig11}
\end{figure}

To investigate details of the band crossing, we examined the $k_x - k_z$ map at the energy of $-1.3$ eV [Fig. \ref{fig11}(a)], $-0.93$ eV [Fig. \ref{fig11}(b)], and $-0.6$ eV [Fig. \ref{fig11}(c)]. 
There are two $k_z$-independent lines at $k_x \sim \pm 0.40 \text{\AA}^{-1}$, which is derived from the TSS above the Dirac point at $-1.75$ eV. 
In addition, one can see a dispersive feature at $k_x \sim \pm 0.5 \text{\AA}^{-1}$ and $k_z \sim 3.0 \text{\AA}^{-1}$ in Fig. \ref{fig11}(a) and $k_x \sim 0 \text{\AA}^{-1}$ and $k_z \sim 2.8 \text{\AA}^{-1}$ in Fig. \ref{fig11}(b). These results indicate the surface and bulk derived states cross around $-1.0$ eV around $k_z = 2.8 \text{\AA}^{-1}$, forming a type-II Dirac point-like band dispersion.

\section{Discussion}
In this study, the TEM/STEM image exhibited the local lattice structure compatible with the CdI$_2$-type with homogeneously mixed Se and Te atoms in the chalcogen layers, which is consistent with the XRD analyses. 
In addition, we have revealed that the bulk and surface-derived energy-band structures in the momentum space are similar to 1T-PdTe$_2$, regardless of the disorder in Se and Te atoms in the chalcogen layers. 
These results may allow us to discuss the increase of the superconducting transition temperature from the viewpoint of the chemical pressure effect.

C. Soulard {\it et al}. examined the lattice parameter and unit cell volume of 1T-PdTe$_2$ as a function of external pressure \cite{soulard2005pressure}. They found that above 15 GPa the intralayer Te-Te distance abruptly expanded and the interlayer Te-Te distance reduced at room temperature \cite{soulard2005pressure}. 
However, the unit cell volume as a function of pressure showed continuous temperature dependence without anomaly. Based on the reduction of the unit cell volume by 10\% for PdSeTe, the chemical pressure is equivalent to an external pressure of ~9 GPa \cite{soulard2005pressure}. 
On the other hand, H. Leng {\it et al.} reported that while $T_{\text{c}}$ enhanced from 1.6 K at ambient pressure up to 1.9 K at 1.0 GPa, $T_{\text{c}}$ reduces above the 1.0 GPa \cite{leng2019superconductivity}. It indicates external pressure cannot increase $T_{\text{c}}$ more than 2 K, exhibiting sharp contrast to the chemical pressure induced by solid solution.
Meanwhile, W. Liu {\it et al.} studied $T_{\text{c}}$ of solid solution PdSe$_{2-x}$Te$_x$ for $x=1.0, 1.1, 1.2$ \cite{liu2021new}. $T_{\text{c}}$ tends to increase with increasing Se concentration ($i.e.,$ chemical pressure) \cite{liu2021new} keeping CdI$_2$-type crystal structure, {\it i.e.}, P$\overline{3}$m1 (No. 164) \cite{liu2021new}. They argued the disorder/defect effects in the enhancement of $T_{\text{c}}$ in addition to the chemical pressure \cite{liu2021new}. While we observed homogeneously mixed Se and Te atoms in the chalcogen layers, the lattice structure as detected by TEM/STEM is rather well defined to be CdI$_2$ type crystal structure. Furthermore, although the ARPES linewidth is somewhat broader due to the disorder/defects scattering, the band dispersion and topological property are still well identified. While we need further { studies} to quantitatively explain the enhancement of $T_{\text{c}}$ based on fine details of the Eliashberg function, { the} present results provide the fundamental { basis for understanding} the topological and superconducting properties of { the} solid solution of PdSeTe.

\section{Conclusion}
We have grown single crystals of 1T-PdSeTe and investigated their crystal structure, electronic structure, and superconducting properties. To the best of our knowledge, for the first time, we have shown that the PdSeTe superconductor possesses topological surface states. The resistivity measurements indicated the superconducting transition temperature ($T_{\text{c}}$) of 3.2 K, which was approximately twice as high as that of 1T-PdTe$_2$ ($T_{\text{c}}$ = 1.64 K). Our magnetic susceptibility measurements revealed a small fraction of surface superconductivity. The scanning transmission electron microscopy showed that Se and Te atoms were homogeneously mixed in the chalcogen layers and that the local symmetry was consistent with the CdI$_2$-type crystal structure. Angle-resolved photoemission spectroscopy results and density functional theory calculations indicated the presence of topological surface states (TSSs) and overall similarity to the bulk-band dispersion of 1T-PdTe$_2$. These results suggest that the TSSs remain robust despite atomic disorder in the chalcogen layers. As the electronic band dispersion and local structures are well-characterized, the enhancement in $T_{\text{c}}$ is likely associated with the chemical pressure. Our findings provide insight into how solid solution influences on surface and bulk electronic states, as well as the superconducting transition temperature.
\\


\section{Acknowledgments}
YK acknowledges Hiroshima University for providing the Graduate School Research Fellowship. YK also thanks Prof. Akio Kimura, Prof. Taichi Okuda, and Prof. Takeshi Matsumura for their valuable discussions and suggestions. YK and KS acknowledge the experimental facilities (XRD and EPMA) at N-BARD, Hiroshima University. The ARPES measurements were performed with the approval of the Proposal Assessing Committee of the Hiroshima Synchrotron Radiation Center (Proposal Number: 22AU002). We thank N-BARD, Hiroshima University, for supplying the liquid helium. This work is partially supported by Grants-in-Aid for Scientific Research (22K03495).


\bibliographystyle{apsrev4-2}
\bibliography{Manuscript_PST_v6}

\end{document}